\documentclass[11pt]{article}

\usepackage{graphicx}
\usepackage[top=.75in,bottom=.75in,left=.75in,right=.75in]{geometry}
\usepackage{amsfonts, amsmath, amssymb, amsthm, constants,comment}
\usepackage[symbol]{footmisc}
\usepackage{bm}
\usepackage{pdflscape}
\usepackage{afterpage}
\usepackage{wrapfig}
\usepackage{makecell}
\usepackage{subcaption}

\usepackage{algorithm,algorithmic}
\usepackage[usenames]{color}
\usepackage[table]{xcolor}  

\usepackage{textgreek}
\definecolor{darkred}{RGB}{100,0,0}
\definecolor{darkgreen}{RGB}{0,100,0}
\definecolor{darkblue}{RGB}{0,0,150}
\definecolor{red}{RGB}{255,0,0}
\definecolor{beaublue}{rgb}{0.74, 0.83, 0.9}
\rowcolors{2}{beaublue!25}{white}

\usepackage{hyperref}
\hypersetup{colorlinks=true, linkcolor=darkred, citecolor=darkgreen, urlcolor=darkblue}
\usepackage{url}

\begin{document}

\title{Permutation invariant multi-output Gaussian Processes for drug combination prediction in cancer }
\author{Leiv Rønneberg\thanks{MRC Biostatistics Unit, University of Cambridge, Cambridge, UK}  \thanks{Oslo Centre for Biostatistics and Epidemiology (OCBE), University of Oslo, Oslo, Norway} \thanks{Corresponding author: lr549 (at) cam ac uk}
\and \rowcolor{white} Vidhi Lalchand \thanks{Broad Institute of MIT \& Harvard}
\and \rowcolor{white} Paul Kirk \footnotemark[1] \thanks{CRUK Cambridge Centre Ovarian Programme}\thanks{Cambridge Institute of Therapeutic Immunology and Infectious Disease (CITIID)} 
}

\date{}
\maketitle

\begin{abstract}
    Dose-response prediction in cancer is an active application field in machine learning. Using large libraries of \textit{in-vitro} drug sensitivity screens, the goal is to develop accurate predictive models that can be used to guide experimental design or inform treatment decisions. Building on previous work that makes use of permutation invariant multi-output Gaussian Processes in the context of dose-response prediction for drug combinations, we develop a variational approximation to these models. The variational approximation enables a more scalable model that provides uncertainty quantification and naturally handles missing data. Furthermore, we propose using a deep generative model to encode the chemical space in a continuous manner, enabling prediction for new drugs and new combinations. We demonstrate the performance of our model in a simple setting using a high-throughput dataset and show that the model is able to efficiently borrow information across outputs.
\end{abstract}

\section*{Introduction}
\label{sec:intro}

Gaussian Processes (GPs) provide a flexible non-parametric \cite{GPBook} paradigm for probabilistic regression and classification. Two of the most prominent properties of GPs are their ability to quantify prediction uncertainty and incorporate inductive biases in function space through the specification of a kernel function. Hand-crafting a kernel function provides a powerful way to incorporate prior knowledge, which we leverage in this work. In this manuscript, we follow up work regarding permutation invariant multi-output Gaussian Processes (PIMOGP) in the context of predicting dose-response functions in drug combination screens in cancer.

\section*{Background}

\subsubsection*{Dose-response prediction}
In recent years, much work has gone in to developing predictive models for cancer treatments. Using data from high-throughput \textit{in-vitro} experiments on multiple drugs and cell lines, large models are trained with the goal of predicting the sensitivity of a drug on a certain cell line. In the context of drug combinations, interest has frequently been on predicting a summary measure of drug interaction, e.g. a synergy score \cite{Menden2019}, computed from fitted dose-response surfaces, and an assumption of how non-interacting drugs should behave. Synergy scores are inherently quite crude measurements of drug interaction, and fundamentally hinge on the choice of non-interaction assumption. For this reason, several authors \cite{comboFM,comboLTR,Ronneberg2023, huusari2024predicting} have proposed algorithms that instead aim to predict the entire dose-response surface. In this paper, which build on our previous work \cite{Ronneberg2023}, we propose a methodology based on multi-output Gaussian Processes (MOGPs). Since dose-response experiments are naturally invariant to the ordering of the drugs in a combination, we further augment the model by directly encoding this invariance into the prior. Inference is performed using stochastic variational inference (SVI), which naturally handles missing data and provides uncertainty quantification.% -- two features that were difficult in our previous implementation.

\subsection*{Gaussian Processes}
A Gaussian Process (GP) is a \textit{stochastic process}, i.e. a collection of random variables, any finite subcollection of which is distributed jointly as a multivariate normal. A GP can be considered a distribution over smooth functions, and is fully characterised by its mean and covariance functions, $\mu(\cdot)$ and $\kappa(\cdot, \cdot)$. We write $f\sim\mathcal{GP}(\mu(\cdot),\kappa(\cdot,\cdot))$ for the GP, and the definition implies that any finite evaluation of the GP at locations $X=\{\mathbf{x}_1,\ldots,\mathbf{x}_n\}$, denoted $\mathbf{f}=[f(\mathbf{x}_1),\ldots,f(\mathbf{x}_n)]^T$ has a multivariate normal distribution $\mathbf{f}\sim\mathcal{N}(\boldsymbol{\mu},K_{ff})$, where the mean vector is constructed by evaluation of the mean function, $\boldsymbol{\mu}=[\mu(\mathbf{x}_1),\ldots,\mu(\mathbf{x}_n)]^T$, and the covariance matrix has entries $\{K_{ff}\}_{ij}=\kappa(\mathbf{x}_i,\mathbf{x}_j)$. For the mean function, it is common to use the zero function, $\mu(\mathbf{x})=0$ for all $\mathbf{x}$, which we also do throughout the paper.

In GP regression, we assume data $\mathcal{D}\{y_i,\mathbf{x}_i\}_{i=1}^n$ is generated from some unknown function $f$ which is given a GP prior:
\begin{align*}
    y_i \vert \mathbf{x}_i, f \ &= f(\mathbf{x}_i) + \epsilon_i \ \ i=1,\ldots,n \\
    f&\sim \mathcal{GP}(0,\kappa(\cdot,\cdot)).
\end{align*}
If the errors are taken as normally distributed, $\epsilon_i\stackrel{iid}{\sim}\mathcal{N}(0,\sigma^2)$, this model is conjugate, and the GP posterior is available in closed form:
\begin{align}\label{eq:GPposterior}
    f \vert \mathbf{y} &\sim \mathcal{GP}(\tilde{\mu}(\cdot),\tilde{\kappa}(\cdot,\cdot)) \nonumber \\
    \tilde{\mu}(\mathbf{x}) &= k(\mathbf{x},X)\left(K_{ff} + \sigma^2\right)^{-1}\mathbf{y} \\
    \tilde{\kappa}(\mathbf{x},\mathbf{x}') &= \kappa(\mathbf{x},\mathbf{x}') - \kappa(\mathbf{x},X)\left(K_{ff}+\sigma^2I\right)^{-1}\kappa(X,\mathbf{x}'), \nonumber
\end{align}
where $\kappa(\mathbf{x},X)=[\kappa(\mathbf{x},\mathbf{x}_1),\ldots,\kappa(\mathbf{x},\mathbf{x}_n)]$ $\kappa(X,\mathbf{x})=\kappa(\mathbf{x},X)^T$, and $\mathbf{y}=[y_1,\ldots,y_n]^T$.

Note that usually, the covariance function depend on some hyperparameters $\boldsymbol{\theta}$ that require careful tuning. This is usually done by maximizing the marginal likelihood, or in the setting of variational inference pursued in this paper some lower bound on the exact marginal likelihood.

\subsubsection*{Sparse GPs}
Naïvely, GPs scale poorly to large datasets -- often viewed as unfeasible above say a couple of thousand observations due to the size of matrices involved in the computations. Evaluating the posterior distribution and the marginal likelihood requires the inversion of the $n\times n$ matrix $(K_{ff}+\sigma^2I)$ which has complexity $\mathcal{O}(n^3)$. In order to deal with this cubic scaling in observations, \textit{sparse GP regression} (SGPR) \cite{snelson2005sparse, titsias2009variational, hensman2013gaussian} has emerged as a simple alternative to exact computations in the context of GPs. SGPR relies on the idea that the full GP posterior can be reasonably well approximated by a smaller set of \textit{inducing variables} $\mathbf{u}\in\mathbb{R}^q$ where $q\ll n$. The inducing variables $\mathbf{u}=[f(\mathbf{z}_1),\ldots,f(\mathbf{z}_q)]^T$ are taken as noiseless evaluations of the GP at some set of \textit{inducing points} $Z=\{\mathbf{z}_1,\ldots,\mathbf{z}_q\}$. 

Following \cite{titsias2009variational} the interest is on approximating the GP posterior in equation (\ref{eq:GPposterior}). For any finite set of function evaluations of $f$, $\mathbf{f}_*$, the posterior predictive can be written as
\begin{align}
    \pi(\mathbf{f}_* \vert \mathbf{y}) = \int \pi(\mathbf{f}_*\vert\mathbf{f})\pi(\mathbf{f}\vert\mathbf{y})\text{d}\mathbf{f},
\end{align}
where $\pi(\mathbf{f}_* \vert \mathbf{f})$ is the conditional GP prior, and $\pi(\mathbf{f}\vert\mathbf{y})$ the posterior distribution over the latent training function values. By employing an extended joint model over the latent variables and the observed data, $\pi(\mathbf{y}\vert\mathbf{f})\pi(\mathbf{f},\mathbf{u},\mathbf{f}*)$ where $\pi(\mathbf{f},\mathbf{u},\mathbf{f}*)$ is the joint prior over the training inputs, the inducing variables, and the new function evaluations, the posterior predictive can be equivalently written as 
\begin{align}
     \pi(\mathbf{f}_* \vert \mathbf{y}) = \int \pi(\mathbf{f}_*\vert\mathbf{f},\mathbf{u})\pi(\mathbf{f}\vert\mathbf{u},\mathbf{y})\pi(\mathbf{u}\vert\mathbf{y})\text{d}\mathbf{f}\text{d}\mathbf{u}.
\end{align}
If the inducing variables $\mathbf{u}$ are chosen optimally, in the sense that $\mathbf{u}$ is a sufficient statistic for $\mathbf{f}$, this equation further simplifies as 
\begin{align}
    \pi(\mathbf{f}_* \vert \mathbf{y}) = \int \pi(\mathbf{f}_*\vert\mathbf{u})\pi(\mathbf{u}\vert\mathbf{y})\text{d}\mathbf{u}.
\end{align}
In practice, $\mathbf{u}$ will not actually function as a sufficient statistic for the latent function evaluations $\mathbf{f}$, and hence the equation above will only be an approximation to the true predictive distribution. Interest is then on optimizing the quality of this approximation, and we introduce a ``free" variational Gaussian distribution $\phi(\mathbf{u}):=\pi(\mathbf{u}\vert\mathbf{f})$, with mean $\boldsymbol{\mu}$ and covariance matrix $\Sigma$. The mean and covariance functions of the approximate GP posterior then becomes
\begin{gather}\label{eq:inducing_mean_covar}
    \mu_{\phi}(\mathbf{x}) = \kappa(\mathbf{x},Z)K_{uu}^{-1}\boldsymbol{\mu} \\
    \kappa_{\phi}(\mathbf{x},\mathbf{x}') = \kappa(\mathbf{x},\mathbf{x}') - \kappa(\mathbf{x},Z)K_{uu}^{-1}\kappa(Z,\mathbf{x}') + \kappa(\mathbf{x},Z)K_{uu}^{-1}\Sigma K_{uu}^{-1}\kappa(Z,\mathbf{x}'), \nonumber
\end{gather}
where $\kappa(\mathbf{x},Z)=[\kappa(\mathbf{x},\mathbf{z}_1),\ldots,\kappa(\mathbf{x},\mathbf{z}_q)]$, $\kappa(Z,\mathbf{x}')=[\kappa(\mathbf{z}_1,\mathbf{x}'),\ldots,\kappa(\mathbf{z}_q,\mathbf{x}')]^T$ and the matrix $K_{uu}$ has entries $\{K_{uu}\}_{ij}=\kappa(\mathbf{z}_i,\mathbf{z}_j)$. These expressions can be efficiently evaluated in $\mathcal{O}(nq^2)$ time, yielding large computational speedups since $q \ll n$. From these equations it is clear that the quality of the approximation will hinge on the parameters of the variational distribution $\phi(\mathbf{u})$ and the inducing locations $Z$. These variational parameters are usually optimized alongside the other parameters of the model by maximizing a lower bound on the marginal likelihood.

Before moving on to the multi-output case, we briefly review variational inference in the context of sparse GPs. Starting from an extended joint model over the data, latent function evaluations and the inducing variables, $\pi(\mathbf{y},\mathbf{u})=\pi(\mathbf{y}\vert\mathbf{f})\pi(\mathbf{f}\vert\mathbf{u})\pi(\mathbf{u})$, we introduce a variational distribution $\phi(\mathbf{u})$ and bound the marginal likelihood using Jensen's inequality:
\begin{align}
    \log \pi(\mathbf{y}) &= \log \int \pi(\mathbf{y}\vert\mathbf{f})\pi(\mathbf{f}\vert\mathbf{u})\pi(\mathbf{u}) \text{d}\mathbf{f}\text{d}\mathbf{u} \nonumber \\
    &\geq \int \pi(\mathbf{f}\vert\mathbf{u})\phi(\mathbf{u})\log \frac{\pi(\mathbf{y}\vert\mathbf{f})\pi(\mathbf{u})}{\phi(\mathbf{u})}\text{d}\mathbf{f}\text{d}\mathbf{u} \\
    &= \int \phi(\mathbf{u})\log \mathcal{N}(\mathbf{y}\vert K_{fu}K_{uu}^{-1}\mathbf{u},\sigma^2I)\text{d}\mathbf{u} - \text{KL}(\phi(\mathbf{u})\Vert \pi(\mathbf{u})) - \frac{1}{2\sigma^2}\text{Tr}(Q), \nonumber
\end{align}
where $Q=K_{ff}-K_{fu}K_{uu}^{-1}K_{uf}$ and $\text{KL}(\phi(\mathbf{u})\Vert \pi(\mathbf{u}))$ denotes the KL-divergence from $\phi(\mathbf{u})$ to $\pi(\mathbf{u})$. Taking the approach of \cite{titsias2009variational}, this bound can be further simplified by integrating out the inducing variables, but we note that this will introduce complex between the observations. Instead, we follow \cite{hensman2013gaussian} and obtain a bound that is separable in the observations:
\begin{align}\label{eq:SVIBound_single}
     \log \pi(\mathbf{y}) &\geq \sum_{i=1}^n \left\{ \log \mathcal{N}(y_i \vert \boldsymbol{\alpha}_i^T\boldsymbol{\mu},\sigma^2) - \frac{1}{2\sigma^2}\boldsymbol{\alpha}_i^T\Sigma\boldsymbol{\alpha}_i-\frac{1}{2\sigma^2}Q_{ii} \right\}- \text{KL}(\phi(\mathbf{u})\Vert \pi(\mathbf{u})),
\end{align}
where $\boldsymbol{\alpha}^T_i=\mathbf{k_i}^TK_{uu}^{-1}$ and $\mathbf{k}_i$ denotes the $i$-th column of $K_{uf}$, $Q_{ii}$ the $i$-th diagonal entry of $Q$, and the variables $\boldsymbol{\mu}$ and $\Sigma$ are from the assumed variational distribution $\phi(\mathbf{u})$. The benefit of this bound is that it allows the use of stochastic variational inference (SVI) with minibatching, which as a bonus will automatically handle missing values in the multi-output setting that are common in dose-response prediction.

\subsubsection*{Multi-output GPs}
Multi-output GPs (MOGPs) are the extension of GP regression to the setting where the regression outputs are multidimensional, i.e. for any input $\mathbf{x}$, the resulting mapping $f(\mathbf{x}) \in \mathbb{R}^m$ for some $m > 1$. There are many ways of constructing MOGPs (see \cite{alvarez2012kernels} for a review), but our focus here will be on the linear model of coregionalisation (LMC).

In the LMC the outputs are modelled as linear combinations of a set of independent latent functions, that are themselves modelled as GPs. That is, considering a set of $m$ outputs $\{f_j(\mathbf{x})\}_{j=1}^m$ for an input $\mathbf{x}\in\mathbb{R}^p$, the $j$-th output is modelled as 
\begin{align}\label{eq:LMC_latent_view}
    f_j(\mathbf{x}) &= a_{j1}u_1(\mathbf{x}) + a_{j2}u_2(\mathbf{x}) + \cdots + a_{jR}u_R(\mathbf{x}),
\end{align}
where $u_r \sim \mathcal{GP}(0,\kappa_r(\cdot,\cdot))$ for $r=1,\ldots,R$ independently, and $a_{j1},\ldots,a_{jR}$ are scalar weights. Note that each latent function is given its own covariance function $\kappa_r$, but that these are free to have the same covariance, while maintaining independence. Latent functions that share their covariance functions can be grouped into $Q$ groups with $R_g$ latent functions in each group, and the equation rewritten as:
\begin{align*}
    f_j(\mathbf{x}) &= \sum_{g=1}^G \sum_{r=1}^{R_g}a_{jg}^{(r)}u_g^{(r)}(\mathbf{x}).
\end{align*}

Since GPs are closed under addition,  this construction induces a GP over all outputs. The cross-covariance between two evaluations $f_j(\mathbf{x})$ and $f_{j'}(\mathbf{x}')$ can be written as
\begin{align*}
    \text{Cov}\left[f_j(\mathbf{x}),f_{j'}(\mathbf{x}')\right] = \sum_{g=1}^G \sum_{r=1}^{R_g}a_{jg}^{(r)}a_{j'g}^{(r)}\kappa_g(\mathbf{x},\mathbf{x}') = \sum_{g=1}^G b_{jj'}^{(g)}\kappa_g(\mathbf{x},\mathbf{x}'),
\end{align*}
where $b_{jj'}^{(g)}=\sum_{r=1}^{R_g}a_{jg}^{(r)}a_{j'g}^{(r)}$. The full cross-covariance over all $m$ outputs can be written as 
\begin{align}
    \mathbf{K}(\mathbf{x},\mathbf{x}') = \sum_{g=1}^G B_g \kappa_g(\mathbf{x},\mathbf{x}'),
\end{align}
where the matrix $B_g$ is known as a \textit{coregionalisation} matrix, with entries $\{B_{g}\}_{ij}=b_{ij}^{(g)}$. In the case of a \textit{complete} dataset where every input is observed at every output, the above expression can be written using Kronecker products:
\begin{align}\label{eq:LMCCovariance}
    \mathbf{K}(X,X) = \sum_{g=1}^G B_g \otimes K_g,
\end{align}
where $K_g$ has entries $\{K_g\}_{ij}=\kappa_g(\mathbf{x}_i,\mathbf{x}_j)$. In the simplest setting, where all latent functions share the same covariance function (i.e. $G=1$), this model is known as the \textit{intrinsic coregionalisation model} (ICM) and the Kronecker structure can be exploited to obtain large computational gains.

For the LMC, the matrices are even larger due to the covariance across outputs, resulting in a cubic complexity in both inputs and outputs -- $\mathcal{O}(n^3m^3)$. In the special case of the IMC, i.e. where $G=1$ in equation (\ref{eq:LMCCovariance}), using various Kronecker tricks can bring this down to $\mathcal{O}(n^3 + m^3)$, which is much more manageable -- see e.g. \cite{SaatchiThesis}. We can leverage SVI in the setting of the LMC as well in order to speed up the necessary computations.

Instead of having a single set of inducing inputs, we allow different inducing locations per latent function $\mathbf{Z}=\{Z_r\}_{r=1}^R$, where $Z_r=\{\mathbf{z}_{1r},\ldots,\mathbf{z}_{qr}\}$, with corresponding observations of each latent function $\mathbf{U}=[\mathbf{u}_1^T,\ldots,\mathbf{u}_R^T]^T$, where $\mathbf{u}_r=[u_r(\mathbf{z}_{1r}),\ldots,u_r(\mathbf{z}_{qr})]^T$. In order to not overload the notation, we assume the same number of inducing variables, $q$, per latent function, and that the dataset is complete, i.e. that every input $\mathbf{x}$ is observed at every output. We denote by $\mathbf{Y}$ and $\mathbf{F}$ vectors of observations and latent evaluations, stacked for each output, i.e. $\mathbf{Y}=[\mathbf{y}_1^T,\ldots,\mathbf{y}_m]^T$ where $\mathbf{y}_j=[y_{1j},\ldots,y_{nj}]^T$, where $y_{ij}=f_{j}(\mathbf{x}_i)+\epsilon_{ij}$, and similarly for $\mathbf{F}$.

The variational bound takes the exact same form as previously, factorising over observations and outputs, but the matrices involved are now more involved, and structured:
\begin{align}\label{eq:SVIBound_MOGP}
     \log \pi(\mathbf{y}) &\geq \sum_{i=1}^n  \sum_{j=1}^m \left\{ \log \mathcal{N}(y_{ij} \vert \boldsymbol{\alpha}_{ij}^T\tilde{\boldsymbol{\mu}},\sigma^2) - \frac{1}{2\sigma^2}\boldsymbol{\alpha}_{ij}^T\tilde{\Sigma}\boldsymbol{\alpha}_{ij}-\frac{1}{2\sigma^2}\tilde{Q}_{ij} \right\}- \text{KL}(\phi(\mathbf{U})\Vert \pi(\mathbf{U})),
\end{align}
where $\boldsymbol{\alpha}_{ij}^T=A_jK_{FU}^{(i)}K_{UU}^{-1}$, $A$ is the matrix with LMC coefficients, i.e. $\{A\}_{ij}=a_{ij}$, where $A_{j}$ denotes the $j$-th row of $A$. The matrix $K_{FU}^{(i)}$ is an $R \times Rq$ block-diagonal matrix with entries $\{k_r(\mathbf{x}_i,Z_r)\}_{r=1}^R$, i.e. formed by evaluating the input point $\mathbf{x}_i$ against all the inducing points $Z_r$ in the $r$-th latent function. $\tilde{Q}_{ij}$ is formed by $A_j[K_{FF}^{(i)}-K_{FU}^{(i)}K_{UU}^{-1}K_{UF}^{(i)}]$, where the matrix $K_{FF}^{(i)}$ is an $R \times R$ block-diagonal matrix with entries $\{k_r(\mathbf{x}_i,\mathbf{x}_i)\}_{r=1}^R$, $K_{UF}^{(i)}=(K_{FU}^{(i)})^T$ and $K_{UU}$ is an $Rq \times Rq$ block-diagonal matrix with entries $\{k_r(Z_r,Z_r)\}_{r=1}^R$. Finally, $\tilde{\boldsymbol{\mu}}$ and $\tilde{\Sigma}$ comes from the assumed variational distribution $\phi(\mathbf{U})$ -- which we will assume a mean-field distribution over $\phi(\mathbf{U})=\prod_{r=1}^R \phi_r(\mathbf{u}_r)$ where $\phi_r(\mathbf{u}_r)=\mathcal{N}(\boldsymbol{\mu}_r,\Sigma_r)$, yielding a block-diagonal $\tilde{\Sigma}$, and a decomposition of the KL-divergence term as $\text{KL}(\phi(\mathbf{U}) \Vert \pi(\mathbf{U}))=\sum_{r=1}^R \text{KL}(\phi_r(\mathbf{u}_r) \Vert \pi_r(\mathbf{u}_r))$.

Prediction at a new input $\mathbf{x}_*$ now amounts to applying equation (\ref{eq:inducing_mean_covar}) for each latent function $u_r(\mathbf{x}_*)$ and linearly combining them for the output of interest according to equation (\ref{eq:LMC_latent_view}).

\subsection*{Permutation Invariance}
In many instances, learning tasks often have known invariances that can be built into models. In image classification for example, images are typically invariant to translations or rotations. In classical neural network architectures, invariance is usually encouraged rather than enforced via \textit{data augmentation} --- including many different version of the same data point with various transformations applied. Working with GPs on the other hand, these invariances can be built directly into the prior through the kernel function \cite{van_der_wilk_learning_2018}.

For the problem of dose-response prediction, interest is on encoding an invariance on the ordering of the drugs --- a \textit{permutation invariance} of the inputs. In the context of multi-output GPs, we want to encode a permutation invariance for every output, $f_j$. Looking at the construction of the LMC in equation (\ref{eq:LMC_latent_view}), it suffices to ensure that each latent function $u_r$ has the required invariance. Letting $\tilde{\mathbf{x}}$ denote a permuted version of the input $\mathbf{x}$, with the desired invariance to encode $u_r(\mathbf{x})=u_r(\tilde{\mathbf{x}})$, this can be achieved by introducing another function $\tilde{u}_r(\mathbf{x})$ and constructing $u_r(\mathbf{x})$ via a summation argument:
\begin{align}\label{eq:invariantsummation}
    u_r(\mathbf{x}) = \tilde{u}_r(\mathbf{x}) + \tilde{u}_r(\tilde{\mathbf{x}}),
\end{align}
from which we see that the mapping $\mathbf{x}\to\tilde{\mathbf{x}}$ leaves the function unchanged. Placing a zero-mean GP prior on $\tilde{u}_r$ with kernel $\tilde{k}_r(\cdot,\cdot)$ induces a zero-mean GP prior on $u_r$ with kernel

\begin{align}
    k_r(\mathbf{x},\mathbf{x}')=\tilde{k}_r(\mathbf{x},\mathbf{x}') + \tilde{k}_r(\mathbf{x},\tilde{\mathbf{x}}') + \tilde{k}_r(\tilde{\mathbf{x}},\mathbf{x}') + \tilde{k}_r(\tilde{\mathbf{x}},\tilde{\mathbf{x}}'). 
\end{align}

In order to enable SVI in the context of permutation invariant MOGPs, a slight modification needs to be made regarding the inducing variables $\mathbf{u}_r$. Specifically, instead of regarding $\mathbf{u}_r$ as observations from the latent function $u_r$, they are assumed as observations from the underlying $\tilde{u}_r$. The bound in equation (\ref{eq:SVIBound_MOGP}) remains unchanged, only requiring some slight changes to some of the matrix entries. Specifically the entries of matrices $K_{FF}^{(i)}$, $K_{FU}^{(i)}$ and $K_{UU}$ need to be computed according to the following equations:

\begin{align}\label{eq:permCovars}
    K_{FF}^{(i)}&: k_{r,FF}(\mathbf{x}_i,\mathbf{x}_i)=\tilde{k}_r(\mathbf{x}_i,\mathbf{x}_i) + \tilde{k}_r(\mathbf{x}_i,\tilde{\mathbf{x}}_i) + \tilde{k}_r(\tilde{\mathbf{x}_i},\mathbf{x}_i) + \tilde{k}_r(\tilde{\mathbf{x}}_i,\tilde{\mathbf{x}}_i) \nonumber \\
    K_{FU}^{(i)}&:k_{r,FU}(\mathbf{x}_i,Z_r) = \tilde{k}_r(\mathbf{x}_i,Z_r) + \tilde{k}_r(\tilde{\mathbf{x}}_i,Z_r) \\
    K_{UU}&: k_{r,UU}(Z_r,Z_r)=\tilde{k}_r(Z_r,Z_r). \nonumber
\end{align}
That is, the entries of these block diagonal matrices are computed according to updated covariance functions $k_{r,FF}(\cdot,\cdot)$, $k_{r,FU}(\cdot,\cdot)$ and $k_{r,UU}(\cdot,\cdot)$ that are themselves functions of different evaluations of the underlying kernel function $\tilde{k}_r(\cdot,\cdot)$ of $\tilde{u}_r$. Prediction at a new input $\mathbf{x}_*$ now amounts to using equation (\ref{eq:inducing_mean_covar}) to produce $u_r(\mathbf{x}_*)$ for each latent $u_r$ (substituting for the updated covariance functions in equation (\ref{eq:permCovars})), and then applying equation (\ref{eq:LMC_latent_view}) for the output of interest.

The model was implemented within the GPyTorch \cite{gpytorch} framework for GP regression.

\section*{Datasets and processing}

\subsection*{Dose-response data}
We use the data from \cite{ONeil2016}, and follow the same pre-processing procedure as in \cite{Ronneberg2023}. This dataset consist of 583 unique combinations of 38 drugs, screened on 39 cell lines across 6 different tissues (Breast, 6; Colon, 8; Lung, 8; Melanoma, 6; Ovarian, 9; Prostate, 2), for a total of 22,737 drug-combination experiments. Within each experiment, drugs are combined and viability read off in a $r \times 4$ grid of concentrations, with 4 replicates at each grid point. Combined with monotherapy experiments performed at anywhere from 6-12 unique concentrations depending on the drug, at around 6 replicates, the dataset totals over 1.2 million viability measurements. 

Each experiment is processed using the bayesynergy \cite{Ronneberg2021} software, which fits a semi-parametrics dose-response function to the data, and provides samples from the posterior predictive dose-response function on a $10 \times 10$ grid of concentrations --- individually scaled to the unit box $[0,1] \times [0,1]$ -- yielding a dataset of 2,273,700 observations on a shared grid of concentrations. As a notable difference from the pre-processing procedure in \cite{Ronneberg2023}, is that instead of training our model on targets derived from the latent GP in the bayesynergy software, we instead take as our targets the fitted values of the dose-response function, which takes values between 0 and 1.

\subsection*{Continuous representation of drugs}
In the PIICM, each drug combination experiment (cell line, drug A, drug B) triplet was considered an output, and only the concentrations $(c_A,c_B)$ was given as inputs. In this manuscript we instead take only the cell line as output, and regard the drugs as inputs alongside the drug concentrations. In order to encode the drug information as inputs, we could one-hot encode them, but instead make use of a deep generative model that takes as input a string representation of the molecule, and outputs a low-dimensional representation of the drug. 

\paragraph{SMILES v. SELFIES} Most of the literature on machine learning based chemical design for the string representation of molecules uses SMILES strings \cite{weininger1988smiles} --- a line notation method which encodes molecular structure using short ASCII strings. However, the SMILES representation has two critical limitations. First, they are not designed to capture molecular similarity, hence molecules with almost identical structure can have markedly different SMILES \cite{jin2018junction}. Second, they are not robust on their own, which means that generative models are likely to produce strings that do not represent valid molecules. Hence, the latent space of DGMs trained on SMILES strings can potentially have large dead zones where none of the points sampled in the region decode to valid molecules. To overcome these issues, we train our model on an alternative string representation for molecules introduced in 2020 \cite{Krenn_2020} that guarantees 100$\%$ robustness --- SELF-referencing embedded string (SELFIES). We do not deep dive into technical construction aspects of the SELFIE syntax in this work, at a high-level one of the difficulties of working with SMILES is the nested bracket closures which appear frequently in the SMILES notation, for instance, consider the smiles string \texttt{CCCc1cc(NC(=O)CN2C(=O)NC3(CCC(C)CC3)C2=O)n(C)n1}, the SELFIE translation uses a  formal Chomsky type-2 grammar or a context-free grammar and gets rid of the non-local characteristics. The molecule above is translated to \texttt{[C][C][C][C][C][=C][Branch2][Ring1][=C][N][C]} \texttt{[=Branch1][C][=O][C][N][C][=Branch1][C][=O]}\texttt{[N][C][Branch1][O][C][C][C][Branch1][C]} \\\texttt{[C][C][C][Ring1][\#Branch1][C][Ring1][N][=O][N]} \texttt{[Branch1][C][C][N][=Ring2][Ring1][\#Branch1]}. We tokenize the SELFIE syntax to represent molecules in our generative model. \\

The deep generative model used in this context is trained for autoencoding and features an open-ended chemical latent space learnt by embedding discrete molecules in a continuous vector space (encoder). For generation, an inverse step (decoder) converts a continuous vector in latent space to a valid molecule. This is the classical encoder-decoder set-up as in a standard VAE \cite{kingma2013auto}. We use a recurrent VAE architecture with an RNN encoder and a decoder to sequentially process the SELFIE representation of the drugs token by token. We embed the cancer drugs in the latent space of the generative model by representing them as SELFIEs and encoding them using the trained encoder. This yields a 50-dimensional latent vector for each drug, i.e. $\mathbf{x}_A\in\mathbb{R}^{50}$ for drug A, for example. Combined with the concentrations, the vector of inputs for each data point becomes $\mathbf{x}=(c_{A},c_{B},\mathbf{x}_A^T,\mathbf{x}_B^T)\in\mathbb{R}^{102}$, with the corresponding permuted version $\tilde{\mathbf{x}}=(c_A,c_B,\mathbf{x}_B^T,\mathbf{x}_A^T)$ that the model is invariant to. Due to some cancer drugs consisting of elements that are rare outside of cancer (such as platinum) that the model was not trained on, only 26 of the 38 drugs were successfully given coordinates in the latent space. Removing experiments for which $\mathbf{x}_A$ or $\mathbf{x}_B$ is missing leaves 289 unique combinations, and a total of 11,271 unique experiments.

%\cite{Lalchand2024}

\subsection*{Prediction setting and hyperparameters}
We test the performance of our model in the \textit{leave-triplet-out} (LTO) setting, using the nomenclature of \cite{abbasi2024new}. That is, we consider prediction of a specific (cell line, drug A, drug B)-triplet that does not appear in the training dataset -- however, the training dataset may contain other examples using the same cell line, or the same drugs. We split the 11,271 experiments 80/20 into a training and test set -- keeping 9016 experiments for training and 2255 for testing.

For 39 cell line outputs, we set the number of latent functions $R=10$ in the LMC all sharing the same kernel function (i.e. G=1 in equation (\ref{eq:LMCCovariance})), and use $q=200$ inducing points for each latent function. For the variational distribution, each component is modelled using a mean-field approximation, $\phi_r(\boldsymbol{\mu}_r,\Sigma_r)$, where $\Sigma_r$ is a diagonal matrix. For the covariance function $\tilde{k}(\mathbf{x},\mathbf{x}')$ we use an RBF kernel over the drug concentrations, and a RBF kernel with automatic relevance determination (ARD) over the drug features. These are then multiplied together to form the final covariance function over the inputs.

We used a batch size of 256, and trained for 12 epochs using the Adam optimizer \cite{kingma2014adam} with an initial learning rate of 1e-2 decreasing to 1e-3 after 6 epochs, and to 1e-4 after 9 epochs. We plot the training loss over the epochs in Figure \ref{fig:train_loss}.

The permutation invariant MOGP (PIMOGP) is compared against the simplest possible model of non-interaction, known as Bliss independence \cite{Bliss1939}. Under this assumption, the dose-response function for a (cell line, drug A, drug B) triplet is simply the product of the two monotherapy functions for drug A and drug B on the cell line. In our experience \cite{Ronneberg2023}, drug interaction is quite rare and Bliss independence is quite often a good approximation to the dose-response surface. As performance metrics, we compute the root mean squared error (RMSE) and Pearson's correlation coefficient

\section*{Results}
\begin{figure}
     \centering
     \begin{subfigure}[t]{0.3\textwidth}
         \centering
         \includegraphics[width=\textwidth]{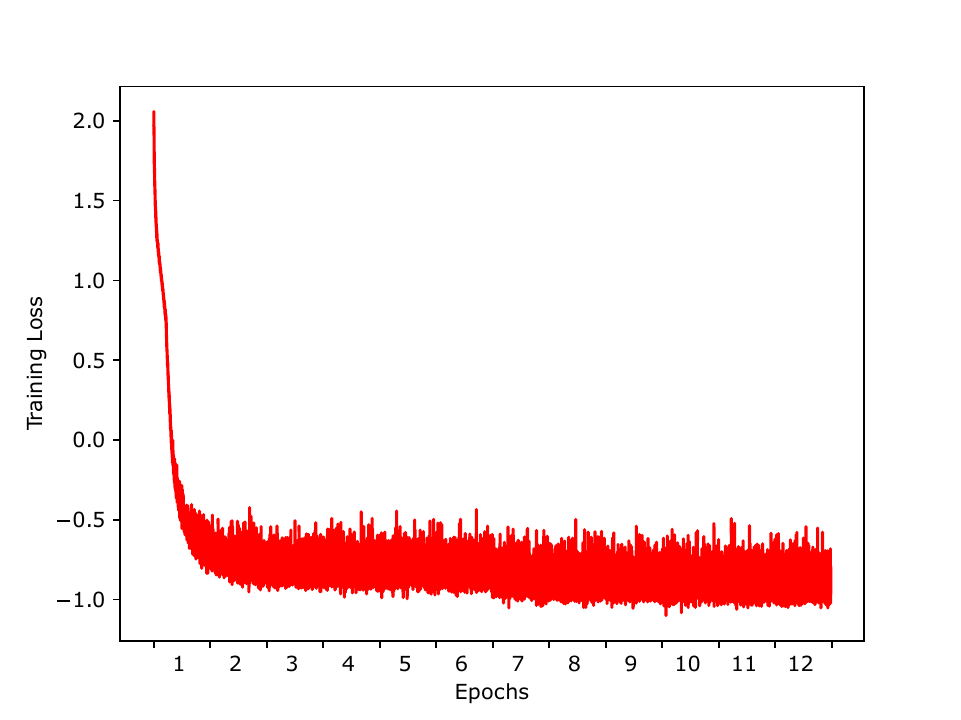}
         \caption{Training loss over 12 epochs}
         \label{fig:train_loss}
     \end{subfigure}
     \hfill
     \begin{subfigure}[t]{0.3\textwidth}
         \centering
         \includegraphics[width=\textwidth]{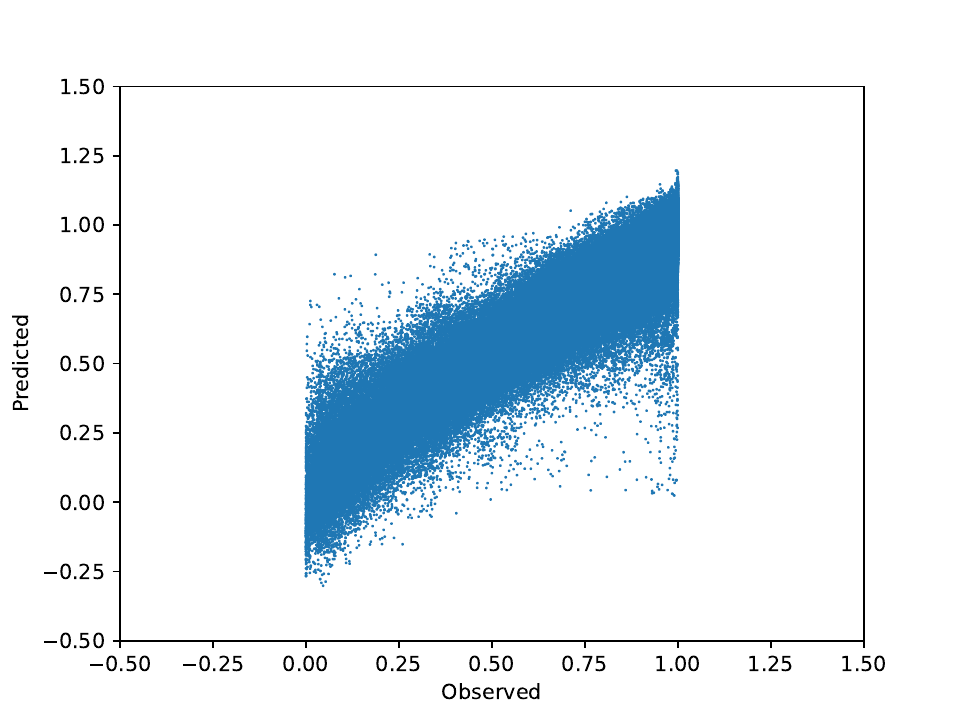}
         \caption{Observed vs. Predicted using the permutation-invariant model}
         \label{fig:obs_pred}
     \end{subfigure}
     \hfill
     \begin{subfigure}[t]{0.3\textwidth}
         \centering
         \includegraphics[width=\textwidth]{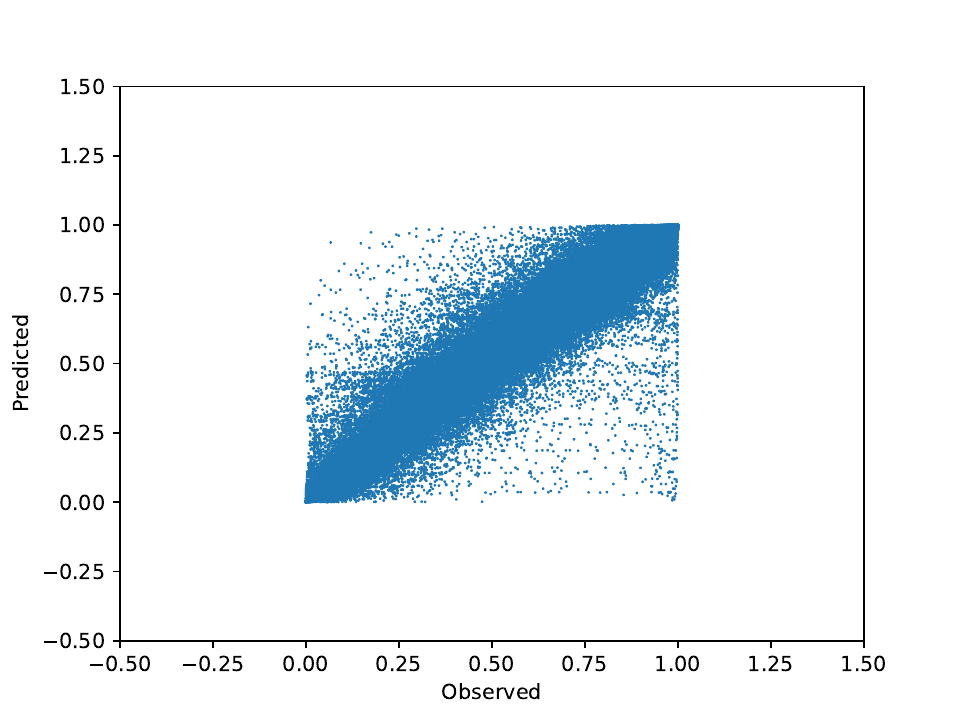}
         \caption{Observed vs. Predicted using the baseline, non-interaction model}
         \label{fig:obs_pred_p0}
     \end{subfigure}
\end{figure}

The results are shown in Table \ref{tbl:Results} and visualised in Figure \ref{fig:obs_pred} and \ref{fig:obs_pred_p0}. We see that our model performs well, reaching a MSE of 0.0956 and a Pearson's r of 0.945. However, in terms of these metrics we are not able to beat the simple baseline model using the Bliss non-interaction assumption, which yields a MSE of 0.0672 and Pearson's r of 0.974. We note though that looking at the figure, the PIMOGP appears to have fewer deviations from the diagonal with large magnitude. These correspond to experiments that are either synergistic or antagonistic, which inherently cannot be captured by the non-interaction model.

\begin{center}
\begin{table}[!ht]
    
    \centering
    \begin{tabular}{l|cccc}
    & \multicolumn{2}{c}{\textbf{PIMOGP}} & \multicolumn{2}{c}{\textbf{Non-interaction}} \\
    Setting & MSE & Pearson's $r$ & MSE & Pearson's $r$ \\
    \hline
    LTO & 0.0956 & 0.945 & \textbf{0.0672} & \textbf{0.974}\\
    \hline
    \end{tabular}
    \caption{Results are shown for PIMOGP and the simple baseline of the non-interaction assumption.}
    \label{tbl:Results}
\end{table}
\end{center}

\begin{figure}
     \centering
     \begin{subfigure}[t]{0.5\textwidth}
         \centering
         \includegraphics[width=\textwidth]{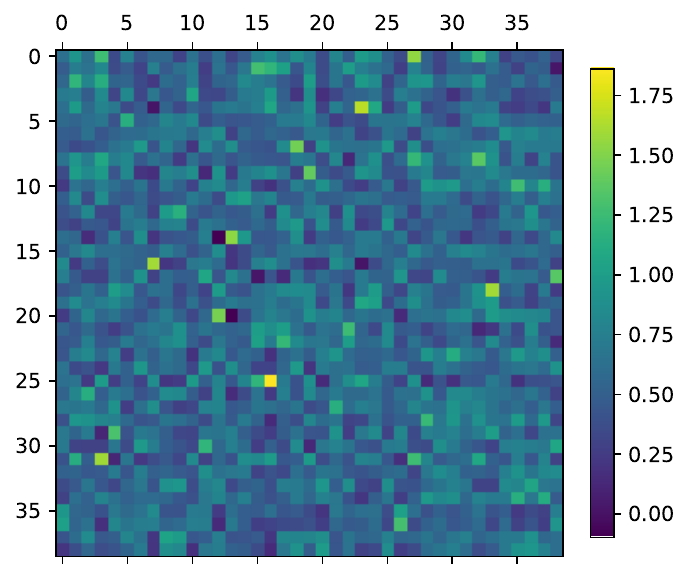}
         \caption{Coregionalisation matrix $B$ giving the learned covariance structure across the 39 cell lines.}
         \label{fig:cell_covar}
     \end{subfigure}
\end{figure}

The performance of the PIMOGP will hinge on the number of latent functions that are used in the LMC, as well as the number of inducing points that are used. By tweaking these, we hope to be able to improve prediction performance.

Finally, in Figure \ref{fig:cell_covar} we plot the coregionalisation matrix $B$ from equation $\ref{eq:LMCCovariance}$, showing the overall covariance across the cell lines. Unlike in \cite{Ronneberg2023}, we see a clear sharing of information across cell lines.

\section*{Conclusion}
In this manuscript, we proposed a multi-output GP method for predicting dose-response surfaces in drug combinations. By carefully constructing a lower bound on the marginal likelihood, and modifying it to incorporate permutation invariance, inference is fast and scalable via stochastic variational inference.

We applied our model to the O'Neil dataset \cite{ONeil2016}, in a simple setting, with a limited number of latent functions and inducing points, and were able to reach a correlation between observed and predicted outputs over 0.94. While we were not yet able to beat the simple (Bliss) baseline for this dataset, we think that by increasing the number of latent functions and number of inducing points this can be achived.

This model formulation is an extension of our previous work \cite{Ronneberg2023} in a number of ways. First, it is based on SVI instead of exact inference, which means that it is more scalable than previously, and naturally handles missing data. Second, in our original formulation, the model could not provide uncertainty quantification due to the sheer size of the matrices involved. With SVI, uncertainy quantification is available with minimal added complexity. Third, by treating drugs as inputs, rather than outputs, and encoding them continuously, the model can in principle be used for other prediction settings, such as the \textit{leave-combination-out} (LCO) or \textit{leave-drug-out} (LDO), to predict dose-response on novel molecules and combinations. Fourth, the model can be further extended to incorporate various \textit{omics} characteristics of the cell lines to predict also for new cell lines. This can be achieved by replacing the co-regionalisation matrix $B$ with a covariance function over e.g. gene expression profiles. We leave this for future work.

\section*{Funding}
This project received funding from the European Union’s Horizon 2020 Research and Innovation Programme under Grant Agreement No. 847912.

\bibliographystyle{unsrt}
\bibliography{reference}

\end{document}

\graphicspath{{figures/}}

\maketitle

\begin{table}[h]
    \centering
    \begingroup
    \setlength{\tabcolsep}{6pt} % Default value: 6pt
    \renewcommand{\arraystretch}{1} % Default value: 1
    \begin{tabular}{llllll}
    \rowcolor{white}
    \multicolumn{6}{c}{\textbf{Supplementary Table 1, cell lines utilized}} \\ 
    \rowcolor{beaublue!50}
    \textbf{Name} & \textbf{Type} & \textbf{Characterizations} & {\colorbox{beaublue!50}{\thead{\textbf{BRAF} \\ \textbf{Status}}}} & {\colorbox{beaublue!50}{\thead{\textbf{p16} \\ \textbf{Status}}}} & {\colorbox{beaublue!50}{\thead{\textbf{p53} \\ \textbf{Status}}}} \\ \hline
    A375	&	melanoma	&	mutations, microarray, RNAseq	&	V600E	&	Mutated	&	WT	\\
    FEMX1	&	melanoma	&	mutations, microarray, RNAseq	&	WT	&	Lacking	&	WT	\\
    FEMXV	&	melanoma	&	mutations, microarray, RNAseq	&	WT	&	Lacking	&	WT	\\
    Hermes 3	&	melanocyte derived	&	mutations, microarray	&	WT	&	Inactive	&	WT	\\
    Hermes 4	&	melanocyte derived	&	mutations, microarray	&	WT	&	Inavtive	&	WT	\\
    HM19	&	melanoma	&		&		&		&		\\
    LOX	&	melanoma	&	mutations, microarray, RNAseq	&	V600E	&	Deletion	&	WT	\\
    MeWo	&	melanoma	&	mutations, microarray, RNAseq	&	WT  &	Mutated	&	Mutated	\\
    Melmet 1	&	melanoma	&	mutations, microarray	&	V600E	&	WT	&	WT	\\
    Melmet 5	&	melanoma	&	mutations, microarray	&	V600E	&	Mutated	&	Mutated	\\
    SKMEL28	&	melanoma	&	mutations, microarray, RNAseq	&	V600E	&	WT	&	Mutated	\\
    WM115	&	melanoma	&	mutations, microarray, RNAseq	&	V600D	&	Hom loss	&	WT	\\
    WM1341	&	melanoma	&	mutations, microarray, RNAseq	&	V600R	&	Deletion	&	WT	\\
    WM1366	&	melanoma	&	mutations, microarray, RNAseq	&	WT	&	Mutated	&	Mutated	\\
    WM1382	&	melanoma	&	mutations, microarray, RNAseq	&	WT	&	WT	&	WT	\\
    WM239	&	melanoma	&	mutations, microarray, RNAseq	&	V600D	&	Hom loss	&	WT	\\
    WM266.4	&	melanoma	&	mutations, microarray, RNAseq	&	V600D	&	Hom loss	&	WT	\\
    WM35	&	melanoma	&	mutations, microarray, RNAseq	&	V600E	&	Deletion	&	WT	\\
    WM45.1	&	melanoma	&	mutations, microarray, RNAseq	&	V600E	&	CN loss	&	Mutated	\\
    WM793b	&	melanoma	&	RNAseq	&		&		&		\\
    WM852	&	melanoma	&	mutations, microarray, RNAseq	&	WT	&	Deletion	&	Mutated	\\
    WM9	&	melanoma	&	mutations, microarray, RNAseq	&	V600E	&	WT	&	WT	\\
    WM983b	&	melanoma	&	mutations, microarray, RNAseq	&	V600E	&	Mutated	&	Mutated	\\
    \end{tabular}
    \caption{Overview of the cell lines utilized in the melanoma screen, alongside the available characterizations and status of BRAF, p16 and p53}
    \label{tbl:celllines}
    \endgroup
\end{table}

\afterpage{%
    \clearpage% Flush earlier floats (otherwise order might not be correct)
\begin{table}[h]
    \centering
    \begingroup
    \setlength{\tabcolsep}{4pt} % Default value: 6pt
    \renewcommand{\arraystretch}{.7} % Default value: 1
    \begin{tabular}{llr}
    \rowcolor{white}
    \multicolumn{3}{c}{\textbf{Supplementary Table 2, drug library}} \\ 
    \rowcolor{beaublue!50}
    \textbf{Name} & \textbf{Type} & \textbf{PubchemID}  \\ \hline
    Abemaciclib	&	Cdk4/6 inhibitor	&	46220502	\\
    all-trans-4-oxo-retinoic acid	&	All-trans retinoic acid	&	444795	\\
    Amsacrine	&	DNA synthesis inhibitor	&	148673	\\
    APO866 (FK866)	&	NAMPT inhibitor	&	6914657	\\
    Avagacestat	&	gamma-secretase (Notch) inhibitor	&	46883536	\\
    Azacitidine	&	Hypomethylating agent	&	9444	\\
    Bardoxolone	&	IKK inhibitor	&	400769	\\
    Bortezomib	&	Proteasome inhibitor	&	387447	\\
    Bosutinib	&	Abl/Src inhibitor	&	5328940	\\
    CCT196969	&	Raf/Src inhibitor	&	42628843	\\
    CHIR-98014	&	GSK3 inhibitor	&	53396311	\\
    Chloroquine	&	Autophagy inhibitor	&	64927	\\
    Cladribine	&	Adenosine deaminase inhibitor	&	20279	\\
    Cytarabine	&	DNA synthesis inhibitor	&	6253	\\
    Daunorubicin	&	Topoisomerase inhibitor	&	30323	\\
    Decitabine	&	Hypomethylating agent	&	451668	\\
    Dexamethasone	&	Corticosteroid	&	5743	\\
    DJ34	&	DNA intercalator and topoisomerase II inhibitor	&	-	\\
    EC-70124	&	Multi-kinase inhibitor	&	44178801	\\
    Enasidenib	&	IDH2 inhibitor	&	89683805	\\
    Entinostat	&	HDAC1/HDAC3 inhibitor	&	4261	\\
    Entospletinib	&	Syk inhibitor	&	59473233	\\
    Everolimus	&	TORC1 inhibitor	&	6442177	\\
    Galunisertib	&	TGF-b inhibitor	&	10090485	\\
    Ganetespib	&	HSP90 inhibitor	&	23624255	\\
    Gilteritinib	&	FLT3/AXL inhibitor	&	49803313	\\
    GSK2879552	&	LSD1 inhibitor	&	66571643	\\
    Ibrutinib	&	BTK inhibitor	&	24821094	\\
    Idelalisib	&	PI3K inhibitor	&	11625818	\\
    IRAK4 Inhibitor (Compound 26)	&	IRAK4 Inhibitor	&   -		\\
    Lapatinib	&	EGFR inhibitor	&	208908	\\
    Lenalidomide	&	Immunomodulatory imide	&	216326	\\
    LY2603618	&	Chk inhibitor	&	11955855	\\
    Methotrexate	&	Antifolate	&	126941	\\
    MI463	&	menin inhibitor	&	90455046	\\
    MK-2206	&	AKT inhibitor	&	46930998	\\
    Mocetinostat	&	HDAC inhibitor	&	9865515	\\
    Molibresib	&	BET inhibitor	&	46943432	\\
    NSC348884	&	Nucleophosmin (NPM1) inhibitor	&	335974	\\
    Olaparib	&	PARP inhibitor	&	23725625	\\
    Panobinostat	&	HDAC inhibitor	&	6918837	\\
    Ponatinib	&	Abl inhibitor	&	24826799	\\
    Quizartinib	&	FLT3 inhibitor	&	24889392	\\
    Ro 5-3335	&	Core binding factor (CBF) inhibitor	&	64983	\\
    Ruxolitinib	&	JAK inhibitor	&	25126798	\\
    Sapatinib	&	Multi-kinase inhibitor	&	11488320	\\
    SB 203580	&	p38MAPK inhibitor	&	176155	\\
    SGC0946	&	Dot1L inhibitor	&	56962337	\\
    SNS-314 Mesylate	&	Aurora kinase inhibitor	&	24995523	\\
    Tacrolimus	&	Calcineurin inhibitor	&	445643	\\
    Tipifarnib	&	Farnesyltransferase inhibitor	&	159324	\\
    Torin2	&	mTOR inhibitor	&	51358113	\\
    Trametinib	&	MEK inhibitor	&	11707110	\\
    Vemurafenib	&	BRAF inhibitor	&	42611257	\\
    Venetoclax	&	Bcl2 inhibitor	&	49846579	\\
    Vismodegib	&	Hedgehog inhibitor	&	24776445	\\
    Volasertib	&	PLK1 inhibitor	&	10461508	\\
    Vorasidenib	&	idh1/2 inhibitor	&	117817422	\\
    WM-8014	&	histone lysine acetyltransferase inhibitor	&	133053564	\\
    WP1130	&	DUB inhibitor	&	11222830	\\
    YM155	&	Survivin inhibitor	&	11178236	\\
    \end{tabular}
    \caption{Overview of the drugs utilized in the melanoma screen.}
    \label{tbl:drugs}
    \endgroup
\end{table}
\clearpage% Flush page
}

\begin{figure}
    \centering
    \includegraphics[width=\textwidth]{figures/cdkn1acrkltp53.pdf}
    \caption{The figure shows the connection between the expression levels of CDKN1A and CRKL, and the mutation status of p53. Mutations in p53 are associated with higher expression of CRKL, and lower expression of CDKN1A.}
    \label{fig:suppfigs1}
\end{figure}

%\newpage

%\bibliographystyle{unsrt}
%\bibliography{reference}